\documentclass[apj,numberedappendix]{emulateapj}
\usepackage{amsmath,amssymb}
\usepackage{array}
\usepackage{academicons}
\usepackage{xcolor}
\usepackage{amsmath}
\usepackage{graphicx}
\usepackage{epstopdf}
\usepackage[colorlinks=true,citecolor=blue,linkcolor=blue,urlcolor=blue]{hyperref}

\usepackage{color}
\definecolor{darkgreen}{rgb}{0.0,0.5,0.0}
\usepackage{array}
\usepackage{booktabs}
\usepackage{multirow}

\begin{document}

\title{Measuring the gravitomagnetic distortion from rotating halos I: methods}
%\adg{suggestion: Measuring the gravitomagnetic distortion from rotating halos in weak lensing surveys I: METHODS}

\author{Chengfeng Tang\altaffilmark{1,2,3}, Pierre Zhang\altaffilmark{1,2,3}, Wentao Luo\altaffilmark{4,1}, Nan Li\altaffilmark{5}, Yi-Fu Cai\altaffilmark{1,2,3}, Shi Pi\altaffilmark{4,6}}

\altaffiltext{1}{Department of Astronomy, School of Physical Sciences, University of Science and Technology of China, Hefei, Anhui 230026, China}
\altaffiltext{2}{CAS Key Laboratory for Research in Galaxies and Cosmology, University of Science and Technology of China, Hefei, Anhui 230026, China}
\altaffiltext{3}{School of Astronomy and Space Science, University of Science and Technology of China, Hefei, Anhui 230026, China}
\altaffiltext{4}{Kavli Institute for the Physics and Mathematics of the Universe (Kavli IPMU,WPI), the University of Tokyo, Chiba 277-8582, Japan}
\altaffiltext{5}{National Astronomical Observatories, Chinese Academy of Sciences 20A Datun Road, Chaoyang District, Beijing, 100012, China}
\altaffiltext{6}{ CAS Key Laboratory of Theoretical Physics, Institute of Theoretical Physics,Chinese Academy of Sciences, Beijing 100190, China }

\email{wentao.luo@ipmu.jp}
\email{yifucai@ustc.edu.cn}
%\email{linan7788626@gmail.com}

\begin{abstract}
%\adg{
Source galaxy images are distorted not only by a static gravitational potential, but also by frame-dragging induced by massive rotating objects like clusters of galaxies.  
Such effect is well understood theoretically, it is therefore of great interest to estimate its detectability for future surveys. 
In this work, we analyze the lensing convergence $\kappa$ around rotating dark matter halos.
The rotation of the massive objects generates a gravitomagnetic potential giving rise to an anisotropic contribution to the lensing potential.
We construct an estimator $\delta \kappa$ to describe the difference between the symmetric enhancement and reduction of $\kappa$ around the halo rotation axis, finding that it is well approximated by a function proportional to the halo velocity dispersion squared times a dimensionless angular momentum parameter. 
Using simulation mocks with realistic noise level for a survey like LSST, we test our estimator, and show that the signal from frame-dragging of stacked rotating lenses is consistent with zero within $1\sigma$.
However, we find that the most massive cluster in SDSS DR7 spectroscopic selected group catalog has a line-of-sight rotation velocity of 195.0km/s and velocity dispersion of 667.8km/s, which is at $1.2\times 10^{-8}$ odds according to the angular momentum probability distribution inferred from N-body simulations. 
By studying SDSS DR7 spectroscopic selected group catalog, we show how rotating clusters can be identified, and, finding that fast rotating clusters might be more abundant than in estimates based on simulations, a detection of gravitomagnetic distortion may be at reach in future surveys. 
\end{abstract}


%\date{\today}

\maketitle

\section{Introduction}\label{sec:introduction}
Gravitational lensing is the phenomenon that light rays from distant galaxies are bent by foreground potentials. 
According to the strength of the distortion, it can be classified as strong gravitational lensing or weak gravitational lensing.

In the strong lensing regime, it is represented by characteristic features of multiple images, or giant arcs. 
This was first observed in 1979~\citep{walsh1979nat} where a quasar (QSO) lensed into two images 0957 + 561 A, B, with the help of a 2.1 meter telescope from Kit Peak National Observatory. Since then, many strong lensing cases have been observed, such as the famous Einstein cross QSO 2237+0305~\citep{huchra1985aj}, the `smiling face' giant arcs SDSS J1038+4849, among 37 systems from the Sloan Digital Sky Survey~\citep{york2000aj,sharon2020}, and SDP 81 from ALAM observation~\citep{hezaveh2016apj}, among others. 
By modeling the position and number of images from strong lensing systems, one can infer the underlined dark matter distribution (see e.g. \cite{tamura2015pasj,caminha2019aa,keeton2001}). 
Time delay measurements from strongly lensed images can be used to put constraints on the Hubble constant $H_0$. \cite{suyu2017mnr,Wong:2019kwg} found  $H_0$ constrained from time delay to be in $4.4\sigma$ tension with measurements from the cosmic microwave background~\citep{planck18} (see also~\cite{birrer2020}, that shows that the tension is alleviated when combining TDCOSMO and SLAC strong lensing catalog as well as taking mass-sheet degeneracy in the modeling). %\cite{sereno2005mnr} pointed out that, the gravitomagnetic effect is to the most 1\% on the constraints of Hubble constant $H_0$, however, as the advent of  wide and deep survey, i.e LSST adn EUCLID \citep{lsst,euclid1},  this 1\% is not negligible any more. For instance, \cite{shajib2018MNRAS} forecastes an $<$1\% constraints on $H_0$ with 40 well measured time delay samples, while several hundreds are going to be dicovered in LSST survey\citep{liao2019ApJ}. So we will explore the bias on $H_0$ brought by this effect in future as well.(Tang et al in prep).

In the weak gravitational lensing regime, the distortion is much smaller, about a few percents of the intrinsic shapes of galaxies.
However,  weak leasing is ubiquitous as long as massive objects are present between distant light sources and the observer. 
%This enables a statistical measure to extract this distortion.  
By stacking multiple images, weak lensing signal can be extracted from spectroscopic surveys.
A commonly used statistics is galaxy-galaxy lensing, which is a powerful tool to study dark matter halos traced by galaxies, or clusters of galaxies (see e.g.~\cite{sheldon2004aj,mandelbaum2006,luo2017apj,luo2018apj}). 
The high order weak lensing statistics, i.e. cosmic shear, can also be obtained from wide-field imaging surveys, such as KiDS~\citep{asgari2020arXiv}, DES~\citep{troxel2018PhRvD}, and HSC-SSP~\citep{hikage2019pasj}. 
%Cosmic shear is sensitive to the amplitude of density perturbations and the fraction of dark matter in the universe. 
Both galaxy-galaxy lensing and cosmic shear, being especially sensitive to the amplitude of density perturbations and the fraction of matter in the universe, can be used to place tight constraints on the cosmological models~\citep{hikage2019pasj, troxel2018PhRvD}. 
In particular, recent studies have shown that weak lensing are useful to constrain various gravity theories, namely,
\cite{chen2019arx} confronted weak lensing observations to a $f(T)$ model from~\cite{cai2020ApJ} (see~\cite{cai2016RPPh} for a review) to test general relativity at galactic scales, while \cite{luo2020} found emergent gravity~\citep{verlinde2017ScPP} to be inconsistent with galaxy-galaxy lensing signals from SDSS DR7 data~\citep{abazajian2009ApJS}.
Given the wealth of information available in the data collected routinely in spectroscopic surveys, it is crucial to scrutinize all potential systematics in order to extract unbiased measurements from gravitational lensing.\\

On the observational side, much efforts are made to mitigate instrumental systematics from e.g. the inaccuracy of the PSF reconstruction~\citep{mandelbaum2005MNRAS,lu2018AJ}, photo-$z$ bias or selection function~\citep{mandelbaum2018MNRAS}.

Another source of potential systematics can arise from the astrophysical properties of the lenses.
In general relativity, a rotating massive object exerts an extra potential through the Einstein-Thirring-Lense effect~\citep{bardeen1975ApJ}. This `frame-dragging' can be seen as a gravitomagnetic distortion in the weak-field approximation. 
This effect is rather subtle for the perihelion of Mercury, contributing to the precession of $-0.002$ arcsec per century~\citep{clemence1947}, which is orders of magnitude smaller than the other sources of precession.
However, it can be important in the vicinity of very-massive objects, e.g. in galaxy clusters~\citep{miller2005AJ, oguri2018}, or galaxy groups~\citep{yang2007apj}. 
At cluster scales, where dark matter halos can reach masses $10^{15}$ times heavier than the sun and fast rotating speed, gravitomagnetic distortion may become significant enough to be measured.
Combined with kinematic Sunyaev-Zel'dovich (SZ) effect~\citep{chluba2002}, this effect can be used to further constrain halo rotation properties, thanks to high-resolution SZ spectral imaging~\citep{mroczkowski2019} .
In weak lensing, rotation of the foreground objects induces additional contribution to the shear (see e.g.~\cite{ciufolini2003,sereno2003,sereno2005mnr,sereno2007}). 
In most analyses, the lenses are assumed to be static, such that the effect caused by the kinematic movement and rotation of foreground objects is neglected. There are also studies of relativisic correction based on Newtonian N-body simulations by applying a nonlinear post-Friedmann framework \citep{bruni2014ApJ,bruni2014PhRvD, Thomas2015MNRAS, milillo2015PhRvD, adamek2016NatPh} or simulations based on f(R) gravity \citep{Thomas2015JCAP}. \cite{bonvin2018redshift}, \cite{gressel2019full} extend such studies to possible observational effects such as redshift distortion and weak gravitational lensing. \cite{Cristian2021MNRAS, barrera2020relativistic} further probe the vector modes and other relativistic effects based on relativistic simulations using GRAMSES code \citep{barrera2020gramses,thomas2015relativistic}. \\

With the advent of large-imaging data from wide-field surveys, such as the Legacy Survey of Space and Time (LSST)~\citep{lsst}, EUCLID~\citep{euclid1}, or the Wide Field Survey Telescope (WFST) \citep{Lou2016SPIE10154E}, it is worthwhile to reassess the detectability of shear distortion produced by gravitomagnetic effect from rotating dark matter halos.
For instance, it has been argued that rotating masses lead to negligible errors in the measurements of the Hubble constant $H_0$ using gravitational time delay~\citep{sereno2005mnr}.
However, \cite{shajib2018MNRAS} forecasts that $H_0$ could be constraint to $<1\%$ with 40 time delay measurements, while it is expected that several hundreds will be detected by LSST~\citep{liao2019ApJ}. 
The impact on the determination of $H_0$ from gravitomagnetic effect will be discussed elsewhere~(Tang et al. in prep.).
%\cite{sereno2005mnr} pointed out that, the gravitomagnetic effect is at most $1\%$ on the constraints of Hubble constant $H_0$, however, as the advent of  wide and deep survey, i.e LSST adn EUCLID \citep{lsst,euclid1},  this 1\% is not negligible any more. For instance, \cite{shajib2018MNRAS} forecasts an $<$1\% constraints on $H_0$ with 40 well measured time delay samples, while several hundreds are going to be dicovered in LSST survey\citep{liao2019ApJ}. So we will explore the bias on $H_0$ brought by this effect in future as well.(Tang et al in prep).
In this paper, we focus on weak lensing.
We investigate to which extent the rotation of halos is relevant in weak lensing measurements from ongoing and future spectroscopic surveys.
We construct a simple estimator, $\delta \kappa$, that measures the anisotropy induced by halo rotation on the lensing convergence field $\kappa$ of stacked clusters, and use simulations to quantify its significance.

The structure of this paper is organized as follows. 
In Sec.~\ref{sec:grm}, we derive the gravitomagnetic distortion induced by rotating lenses with a singular isothermal sphere (SIS) density profile, and construct an estimator $\delta\kappa$ to quantify the resulting anisotropic contribution to the convergence field $\kappa$. 
Simple approximation for $\delta\kappa$ are given as a function of the velocity dispersion, angular momentum, and rotation axis orientation, of the lens.
In Sec.~\ref{sec:toy}, using GalSim~\citep{rowe2015} and astropy4.0.1~\footnote{\url{https://www.astropy.org/}}, we generate a set of simulation mocks for LSST-like surveys of the lensing anisotropic signal from rotating halos, with various choices for the model parameters. 
In Sec.~\ref{sec:result}, we analyze the signal-to-noise ratio of $\delta \kappa$ given the various configurations of the simulations, and discuss using SDSS DR7 group catalog~\citep{yang2007apj} the characteristics of rotating galaxy clusters and their identification.
We conclude in Sec.~\ref{sec:conclusion}.

The code developed to measure gravitomagnetic effects in $\kappa$ maps is made publicly available via Zenodo \cite{luo_wentao_2021_4445484}. 

%The structure of this paper is organized as follows. In Sec.~\ref{sec:grm}, we present the mathematical derivation of gravitomagnetic distortion based on singular isothermal sphere (SIS) density profile, and we create the concept of $\delta\kappa$, an estimator to quantify this effect using $\kappa$ field based on weak lensing. We fit a functional relation of $\delta\kappa$ as a function of both velocity dispersion and rotation parameter of halos $\lambda$, as well as the azimuthal dependence with respect to the rotation axis. In Sec.~\ref{sec:toy}, we code up the maths in Sec.~\ref{sec:grm} with PYTHON. GalSim \cite{rowe2015} and astropy4.0.1 under BSD license to simulate set of $\kappa$ field with different parameters, i.e. velocity dispersion, rotation parameter $\lambda$ and the orientation of the axis $\psi$. We also add scatters for $\psi$ to test how this scatter influence the $\delta\kappa$ estimation. The results are discussed in Sec.~\ref{sec:result} and we conclude in Sec.~\ref{sec:conclusion}.
%Throughout this paper, we stick to PLANCK18 cosmology \cite{planck18} with $\Omega_m=0.315$, $\Omega_{\Lambda}=0.685$, $\sigma_8=0.811$ and $H_0=67.4$. The unit of angles in this paper is all in degree, or we will specify in the contents.

\section{Gravitomagnetic effect}
\label{sec:grm}
In this section, we derive the gravitomagnetic distortion induced by rotating halos with a Singular Isothermal Sphere (SIS) density profile.

\subsection{$\kappa$ field from gravitomagnetic effect}
We work within the parameterized post-Newtonian (PPN) approximation and consider dark matter halo with a SIS density profile. 
For static halos, the lensing convergence $\kappa(\pmb{\xi})$ reads~\citep{bartelmannmain2001}:
\begin{equation}
    \kappa(\pmb{\xi})=\frac{\Sigma(\pmb{\xi})}{\Sigma_{crit}(z_l,z_s)}=\frac{\sigma_v^2}{2G \, \Sigma_{crit}(z_l,z_s)\, \pmb{\xi}},
\end{equation}
where $\Sigma_{crit}(z_l,z_s)=\frac{c^2}{4\pi G}\frac{D_s}{D_lD_{ls}}$ is the geometry factor given a lens at redshift $z_l$ and a source at $z_s$, $D_l$, $D_s$ and $D_{ls}$ are the angular diameter distances between the observer and the lens, the observer and the source, and the lens and the source, respectively, $\sigma_v^2$ is the velocity dispersion of matter in the halo, and the vector $\pmb{\xi}$ denotes the two-dimensional position with respect of the gravitational potential center of the dark matter halo. 

\begin{figure}
    \centering
    \includegraphics[width=8cm,height=6cm]{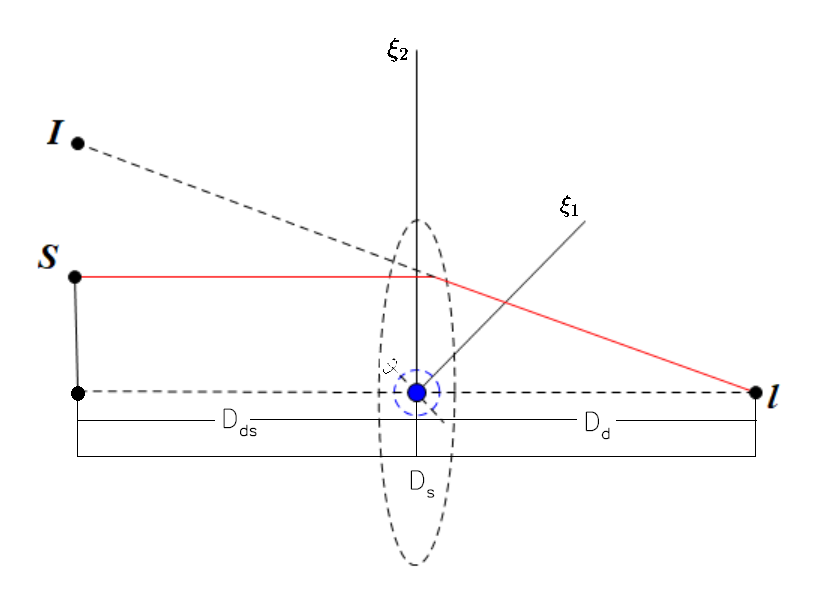}
    \caption{The diagrammatic sketch of a gravitational lens system. The light ray propagates from the source $S$ from the optic axis to the observer $l$. The lens plane is orthogonal to the line of sight, which are shown as $\xi_1-\xi_2$ plane. The distances between the observer and the source, the observer and the lens, and the lens and the source are $D_{s}, D_{d},$ and $D_{ds},$ respectively.}
    \label{fig:1}
\end{figure}
%The shear field is then can be calculated from kappa field\citep{bartelmannmain2001}
%\begin{equation}
%\label{eq:shear}
    %\mathcal{\gamma}(\pmb{\xi})=\frac{1}{\pi}\int_{\mathcal{R}^2}d^2\xi^2\mathcal{D}(\pmb{\xi}-\pmb{\xi^{\prime}})\kappa({\pmb{\xi^{\prime}}}),
%\end{equation}
%where $\mathcal{D}(\pmb{\xi}-\pmb{\xi^{\prime}})$ is the kernel that
%\begin{equation}
    % \mathcal{D}(\pmb{\xi})=\frac{\xi_2^1-\xi_1^2-2i\xi_1\xi_2}{|\xi|^4}.
%\end{equation}
%In the case of SIS density profile, The 2 components of Eq.\ref{eq:shear} can be written by:
%\begin{gather} 
%\gamma_1(\pmb{\xi}) = %\kappa(\pmb{\xi})cos[2\phi(\xi)] ,\\ 
%\gamma_2(\pmb{\xi}) = %\kappa(\pmb{\xi})sin[2\phi(\xi)]. 
%\end{gather}
%where 
%\begin{gather} 
%cos[2\phi(\xi)] =  \frac{\xi_2^2-\xi_1^2}{|\xi|^2}, \\ 
%sin[2\phi(\xi)] =  \frac{2\xi_1\xi_2}{|\xi|^2}.
%\end{gather}
%
Let us now turn to rotating halos. 
The halo angular momentum $J$ can be quantified by a dimensionless parameter $\lambda$, the ratio between the actual angular velocity and the theoretical one~\citep{padmanabhan2002theoretical}:
\begin{equation}
    J=\lambda\frac{GM^{5/2}}{|E|^2},
\end{equation}
where $M$ and  $E$ are the total mass and the total energy of the halo. 
The parameter $\lambda$ is almost independent of halo mass and of the large-scale structure.
Its distribution is approximated by a log-normal function~\citep{vitvitska2002ApJ}:
\begin{equation}
\label{eq:plambda}
    p(\lambda)d\lambda=\frac{1}{\sqrt{2\pi}\sigma_{\lambda}}\exp\big[-\frac{\ln^2(\lambda/\bar{\lambda})}{2\sigma_{\lambda}^2}\big]\frac{d\lambda}{\lambda},
\end{equation}
where the mean value $\bar{\lambda}\approx 0.05$ and the scatter $\sigma_{\lambda}$ is around 0.5. 
For halos with SIS profile, the angular momentum and energy are  related to the total mass $M_{SIS}=\frac{2\sigma_V^2}{G}R_{SIS}$ as~\citep{bartelmannmain2001}:
 \begin{align}
&E_{SIS}=-M_{SIS}\sigma_v^2,
\\
&J_{SIS}=\lambda\frac{4\sigma_v^3R_{SIS}^2}{G},
 \end{align}
where $R_{SIS}\gg|\xi|$ is the truncation radius. 
Following~\cite{sereno2005mnr}, we take the truncated radius such that the mean density within the radius is $\sim 200$ times larger than the critical density, yielding:
\begin{equation}
\label{eq:trunc_radius}
    R_{SIS}=\frac{2\sigma_v}{\sqrt{n}H(z)},
\end{equation}
where $n\sim 200$ characterizes the density ratio between the halo region and the mean density of the universe, and $H(z)$ is the Hubble parameter at redshift $z$.
%Here we introduce a length-scale: $R_{E}}=4 \pi\left(\frac{\sigma_{v}}}{c}\right)^{2} \frac{D_{d}} D_{ds}}}{D_{s}}}$. $R_{sis}$ can be expressed by $R_{E}}$. In this paper, we typically choose $R_{sis}=1000R_{E}}$. 
Although the halo velocity depends on multiple physical processes such as merging, we assume for simplicity that the rotation pattern is stable within the observational time.
Under these assumptions, the halo rotation adds an extra term to the lensing potential via the Einstein-Thirring-Lense effect on top of the SIS potential~\citep{sereno2005mnr}:
\begin{equation}
  \phi=\phi_0+\phi_{GRM},
\end{equation}
where the spherical SIS halo lensing potential reads:
\begin{equation}
\phi_0(\pmb{\xi})=\frac{4G}{c^2}\int_{\mathcal{R}}d^2\pmb{\xi'}\Sigma(\pmb{\xi'}) \ln|\pmb{\xi}-\pmb{\xi'}|=\frac{D_{ds}}{D_s}\frac{4\pi\sigma^2}{c^2}|\xi| \, .
\end{equation}
The extra potential term introduced by gravitomagnetic effect $\phi_{\rm GRM}$ by a SIS density profile with velocity $\pmb{v}$ is given by~\citep{sereno2005mnr}:
\begin{equation}
    \phi_0(\pmb{\xi})=\frac{4G}{c^2}\int_{\mathcal{R}}d^2\pmb{\xi'}\Sigma(\pmb{\xi'})\langle \pmb{v}\cdot \pmb{e}_{l}\rangle \ln|\pmb{\xi}-\pmb{\xi'}|.
\end{equation}
where $\langle \pmb{v}\cdot\pmb{e}_l \rangle$ is the average velocity along the line of sight weighted by the projected density:
\begin{equation}
   \langle\pmb{v}\cdot\pmb{e}_l\rangle(\pmb{\xi}) = \frac{\int dl\big[ \pmb{v}(\pmb{\xi'},l)\big]\cdot \pmb{e}_l\rho(\pmb{\xi},l)}{\Sigma(\pmb{\xi})}.
\end{equation}
The gravitomagnetic deflection angle is then given by the derivative of~$\phi_{GRM}$, yielding:
\begin{equation}
\label{eq:angleGRM}
    \hat{\pmb{\alpha}}_{GRM}(\pmb{\xi})= -\frac{8G}{c^2}\int_{\mathcal{R}^2}d^2\xi^{'}\Sigma(\pmb{\xi}^{'})\frac{\langle \pmb{v}\cdot\pmb{e}_l\rangle(\pmb{\xi^{'}})}{c}\frac{\pmb{\xi}-\pmb{\xi^{'}}}{|\pmb{\xi}-\pmb{\xi^{'}}|^2}.
\end{equation}

\begin{figure}
    \centering
    \includegraphics[width=5cm,height=5cm]{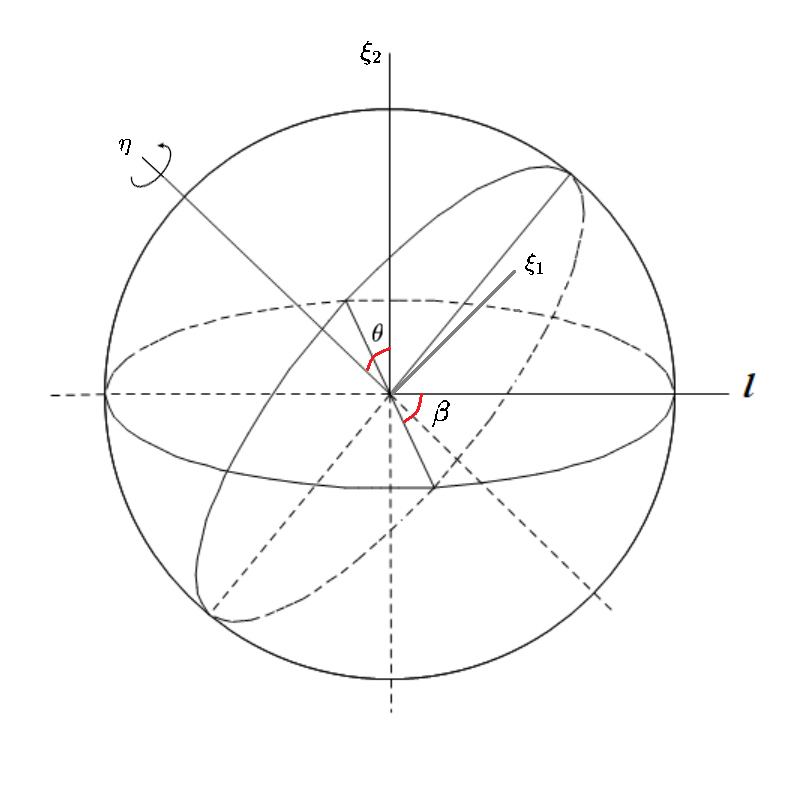}
    \caption{Geometric relationships among the line of sight $l$, the rotation axis $\hat{\eta}$, the lens planes $\xi_1$ and $\xi_2$, and the two Euler angles $\theta$ and $\beta$.}
    \label{fig:2}
\end{figure}

Let us now consider a spherically symmetric lens that rotates about an arbitrary axis $\hat{\eta}$, passing through its center (i.e. a main axis of inertia). To specify the
orientation of the rotation axis, we need two Euler angles: $\theta$, the angle between $\hat{\eta}$ and the $\xi_2$ axis, and $\beta$, the angle between the line of sight $l$ and the line of nodes defined at the intersection of the $\xi_1$ plane and the equatorial plane (i.e., the plane orthogonal to the rotation axis and containing the lens center). 
The sketch map is shown from Fig.~\ref{fig:1} and Fig.~\ref{fig:2}. The former illustrates the geometry of the lensing system at large scales, while the latter, zoomed in, describes the relations among the line of sight, rotation axis, etc.
Under the axial symmetry about the rotation axis, we get:
\begin{align}
\label{eq:vlos1}
   \langle \pmb{v}\cdot\pmb{e}_l \rangle(\xi_1,\xi_2,l) & =-\omega(R)\big[\xi_1 \cos(\theta)+\xi_2 \sin(\theta)\cos(\beta)\big]\notag \\
   & =-\omega_1(R)\xi_1+\omega_2(R)\xi_2 \, ,
\end{align}
where $\omega(R)$ is the modulus of the angular velocity at a distance $R=\sqrt{R_1^2+R_2^2}$ from the rotation axis. 
In the case of a rigid body, there is no dependence on $R$ for $\omega$, and Eq.~\eqref{eq:vlos1} simplifies to: $\pmb{v}\cdot\pmb{e}_l(\xi_1,\xi_2,l)=-\omega_1(R)\xi_1+\omega_2(R)\xi_2.$
\begin{figure*}
    \centering
    \includegraphics[width=8cm,height=6cm]{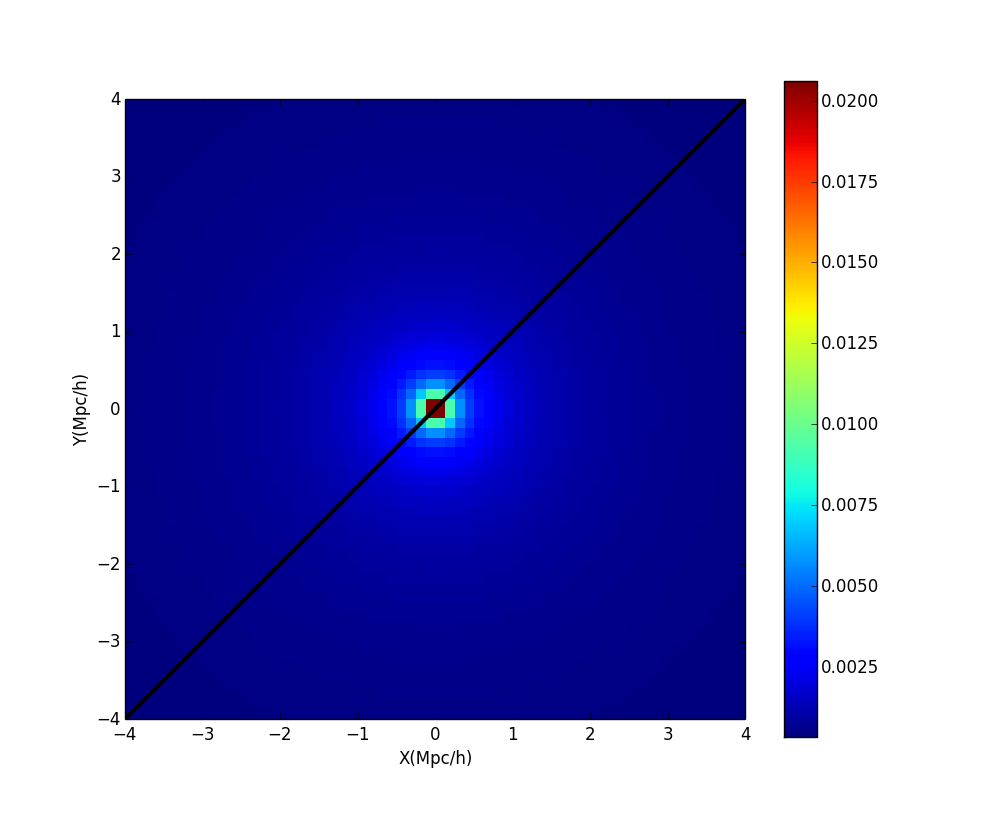}
    \includegraphics[width=8cm,height=6cm]{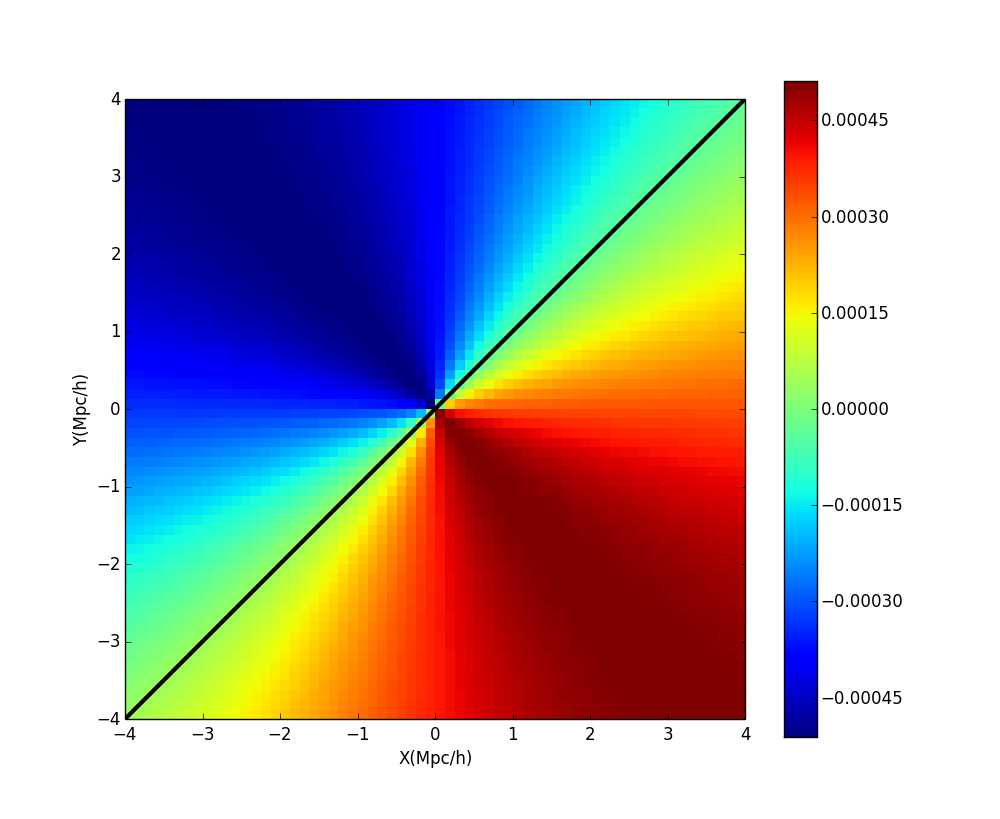}
    \caption{$\kappa$ map distribution of a rotational SIS halo, the black line indicates the rotation axis. The left panel is for a static case, the right one is the dipole from the gravitomagnetic potential induced by rotation. We show the static $\kappa$ map and the rotation $\kappa$ separately due to the fact that the latter is two to three orders magnitude lower than the static signal, even when we set the $\lambda$ value to 0.99 which is close to the limit for a bounded system.}
    \label{fig:kappa}
\end{figure*}
Thus, for spherically-symmetric rigid halos, the gravitomagnetic deflection angle, Eq.~\eqref{eq:angleGRM}, becomes:
\begin{equation}
\label{eq:angleGRM1}
   \begin{array}{l}
\hat{\alpha}_{1}(\pmb{\xi})_{GRM}=\frac{2\kappa}{3c}\left[\frac{\omega_{2}\left(2 \xi_{1}^{2}+\xi_{2}^{2}\right)-\omega_{1} \xi_{1} \xi_{2}}{|\xi|}-\frac{3 \omega_{2}R_{SIS}}{2 }\right], \\ \\
\hat{\alpha}_{2}(\pmb{\xi})_{GRM}=\frac{2\kappa}{3c}\left[\frac{\omega_{1}\left(2 \xi_{2}^{2}+\xi_{1}^{2}\right)-\omega_{2} \xi_{1} \xi_{2}}{|\xi|}-\frac{3 \omega_{1}R_{SIS}}{2 }\right],
\end{array}
\end{equation}
where $\xi=\sqrt{\xi_1^2+\xi_2^2}$.

By taking the derivative of Eq.~\eqref{eq:angleGRM1}, we obtain the contribution from gravitomagnetic effect to the $\kappa$ field:
\begin{equation}
\label{eq:kappa_grm}
\kappa(\pmb{\xi})_{GRM}=\frac{D_d}{2}(\nabla_{\xi} \cdot \hat{\pmb{\alpha}})=\frac{\kappa(\pmb{\xi})\left(\omega_{2} \xi_{1}-\omega_{1} \xi_{2}\right)}{c}.
\end{equation} 

\subsection{$\delta\kappa$ estimator}
Gravitomagnetic effect induced by rotating lens yields an anisotropic contribution to the lensing convergence field $\kappa$ in the form of a dipole, as given by Eq.~\eqref{eq:kappa_grm}.
This is illustrated in Fig.~\ref{fig:kappa}. 
The first panel shows the $\kappa$ field for a static SIS halo, with velocity dispersion of $1000$km/s. 
The second panel depicts the contribution by gravitomagnetic effect with $\lambda=0.2$. 
It shows an enhancement from one side of the rotation axis and reduction from the other side. 
It is clear from Fig.~\ref{fig:kappa} that the dipole induced by the rotation of the lens is much smaller than the total signal. We can however be a bit more quantitative in order to investigate if such gravitomagnetic distortion can be detected.

One can construct an estimator to quantify the anisotropy of the signal by taking the difference of the mean $\kappa$ divided by the rotation axis:
\begin{equation}
\label{eq:deltakappa}
\delta\kappa=\langle \kappa_{enhance}\rangle-\langle \kappa_{reduce}\rangle.
\end{equation}
The mean value of $\kappa$ from both side of the rotation axis is measured inside the virial radius as in Eq.~\eqref{eq:trunc_radius} of a halo to minimize the effect from large-scale structure. The virial radius of a $10^{14}$ solar mass is about 2Mpc/h according to this calculation, which is similar to that of an NFW profile~\citep{nfw1997ApJ}, based on numerical simulations. 

\begin{figure}
   %\centering
    \includegraphics[width=8cm,height=6cm]{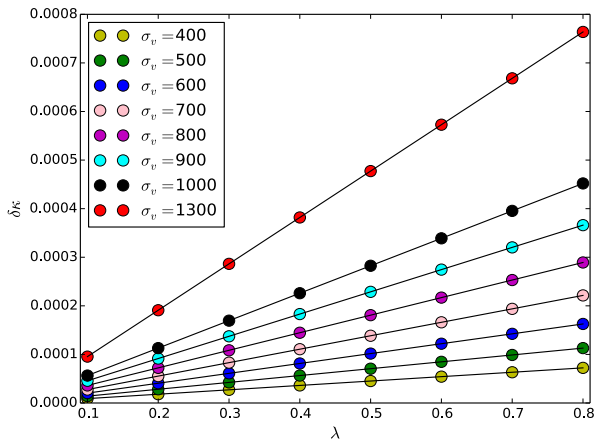}
    \caption{$\delta\kappa$ as a function of the fractional angular momentum parameter $\lambda$ and the velocity dispersion of dark matter inside halos $\sigma_v$ (in [km/s]).}
    \label{fig:lamdk}
\end{figure}

In the upper right panel of Fig.~\ref{fig:kappa}, the two regions can be clearly seen in two different colors. 
In Fig.~\ref{fig:lamdk}, we evaluate the dependence of $\delta\kappa$ on the halo velocity dispersion $\sigma_v$ and the rotation parameter $\lambda$.
To do so, we generate $\kappa$ maps induced by gravitomagnetic effects given by Eq.~\eqref{eq:kappa_grm} and measure $\delta\kappa$ as described above.
We see that $\delta\kappa$ gets bigger as $\sigma_v$ or $\lambda$ gets bigger. 
For a halo with velocity dispersion of $1300$km/s and rotation parameter of $0.8$, $\delta\kappa$ can be comparable in size to $\kappa$ at the edge of the halo.

We now look for a simple functional form for $\delta\kappa$.
As for a SIS profile, there is no rotation if there is no velocity dispersion, $\delta\kappa$ depends at leading order on $\sigma_v^2$.
We measure $\delta\kappa$ on a grid of $500 \times 500$ points of ($\lambda$, $\sigma_v$) and fit with the following ansatz:
\begin{equation}
\label{eq:dkap}
   \delta\kappa(\lambda,\sigma_v)=\mu \, \lambda \, \left( \frac{\sigma_v}{1000 {\rm km/s}} \right)^2 ,
\end{equation}
where we get $\mu \sim 5.69 \cdot 10^{-4}$, an overall normalization that can be considered as the typical magnitude of gravitomagnetic effects from rotating halos. 
Eq.~\eqref{eq:dkap} shows that there is a strong degeneracy between $\sigma_v^2$ and $\lambda$: they are thus strongly dependent on observations. 

We can also express Eq.~\eqref{eq:dkap} as a function of the halo mass:
\begin{equation}
%     \delta\kappa(\lambda,\sigma_v)=(\mu GM/2R_{SIS})\lambda,
 \delta\kappa(\lambda,\sigma_v)=\mu \, \lambda \,  \frac{GM}{2R_{SIS}},
\end{equation}
where $R_{SIS}$ is the truncation radius defined in Eq.~\eqref{eq:trunc_radius}.
Taking into account the distribution of $\lambda$ as given in Eq.~\eqref{eq:plambda}, Eq.~\eqref{eq:dkap} becomes:
\begin{equation}
\label{eq:dkap1}
    \langle \delta\kappa(\sigma_v)\rangle=\int P(\lambda)(\mu\sigma_v^2)\lambda d\lambda \, .
\end{equation}
%Later we will see if velocity dispersion
%can be accurately measured, then $\lambda$ can be recovered to an accuracy of 10\% given a single massive cluster of $\sigma_v=1000$km/s, $\lambda=0.2$, shape noise level 0.3 and galaxy number density of 50 per square arcminute.
So far we have considered that the real (halo) rotation axis is perfectly aligned with the observed (tracer) rotation axis.
If there is a misalignment $\delta\theta$ between those two, $\delta\kappa$ will be reduced by a factor of $\cos(\delta\theta)$, such that the misaligned $\delta\kappa_{m}$ is related to the aligned $\delta\kappa_{o}$ by:
\begin{equation}
\label{eq:dkapm}
    \delta\kappa_{m}=\cos(\delta\theta)\delta\kappa_{o} .
\end{equation}
For a given distribution $P(\delta\theta)$, Eq.~\eqref{eq:dkapm} becomes:
\begin{equation}
\label{eq:pdelta}
    \langle \delta\kappa \rangle=\int d\delta\theta P(\delta\theta) \cos(\delta\theta)\delta\kappa \, .
\end{equation}
Combining Eq.~\eqref{eq:dkap},  Eq.~\eqref{eq:dkapm} and Eq.~\eqref{eq:pdelta}, 
%we obtain the general form of $\delta\kappa$ to mimic the real estimator with both scatter on $\lambda$ and misalignment effect. which is
we obtain a general expression for $\delta\kappa$ taking into account the scatters of both the angular momentum and the misalignement of the observed-to-real rotation axis: 
\begin{equation}
\label{eq:dkapall}
        \langle \delta\kappa \rangle=\int d\delta\theta P(\delta\theta) \cos(\delta\theta)\int P(\lambda)\mu\lambda(\frac{\sigma_v}{1000km/s})^2 d\lambda.
\end{equation}

To summarize, 
Eq.~\eqref{eq:dkapm} describes gravitomagnetic distortion around a single cluster, while Eq.~\eqref{eq:dkapall} provides an estimator for stacked clusters.
We will discuss next this later case, asking ourselves if stacking multiple rotating lenses can achieve sufficient signal-to-noise ratio (SNR) for a measurements of gravitomagnetic effect. 
Meanwhile, we finish this section by discussing related issues on the measurements of gravitomagnetic distortion in the single cluster case. 

\begin{figure}
   %\centering
    \includegraphics[width=9cm,height=7cm]{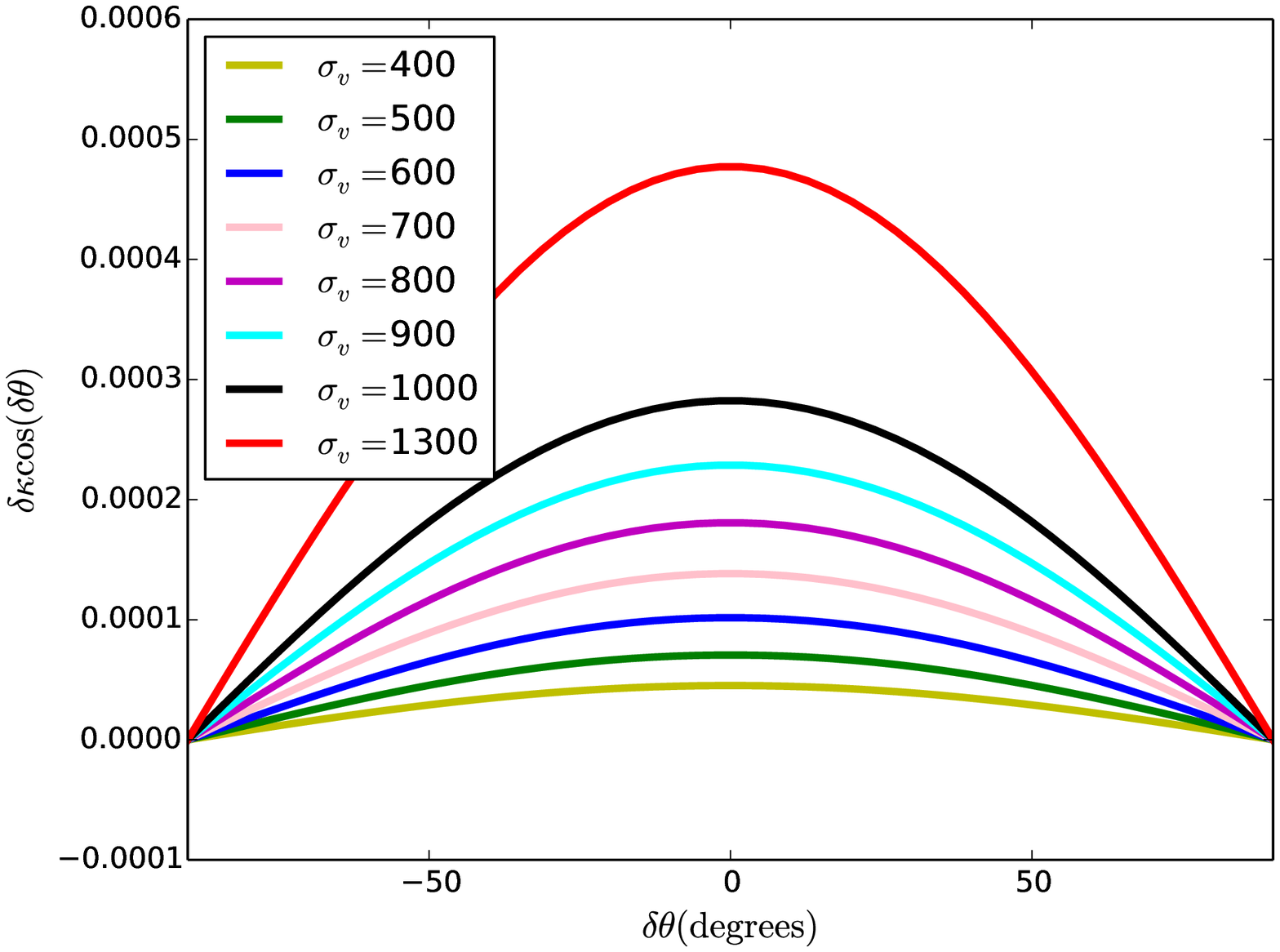}
    \includegraphics[width=9cm,height=7cm]{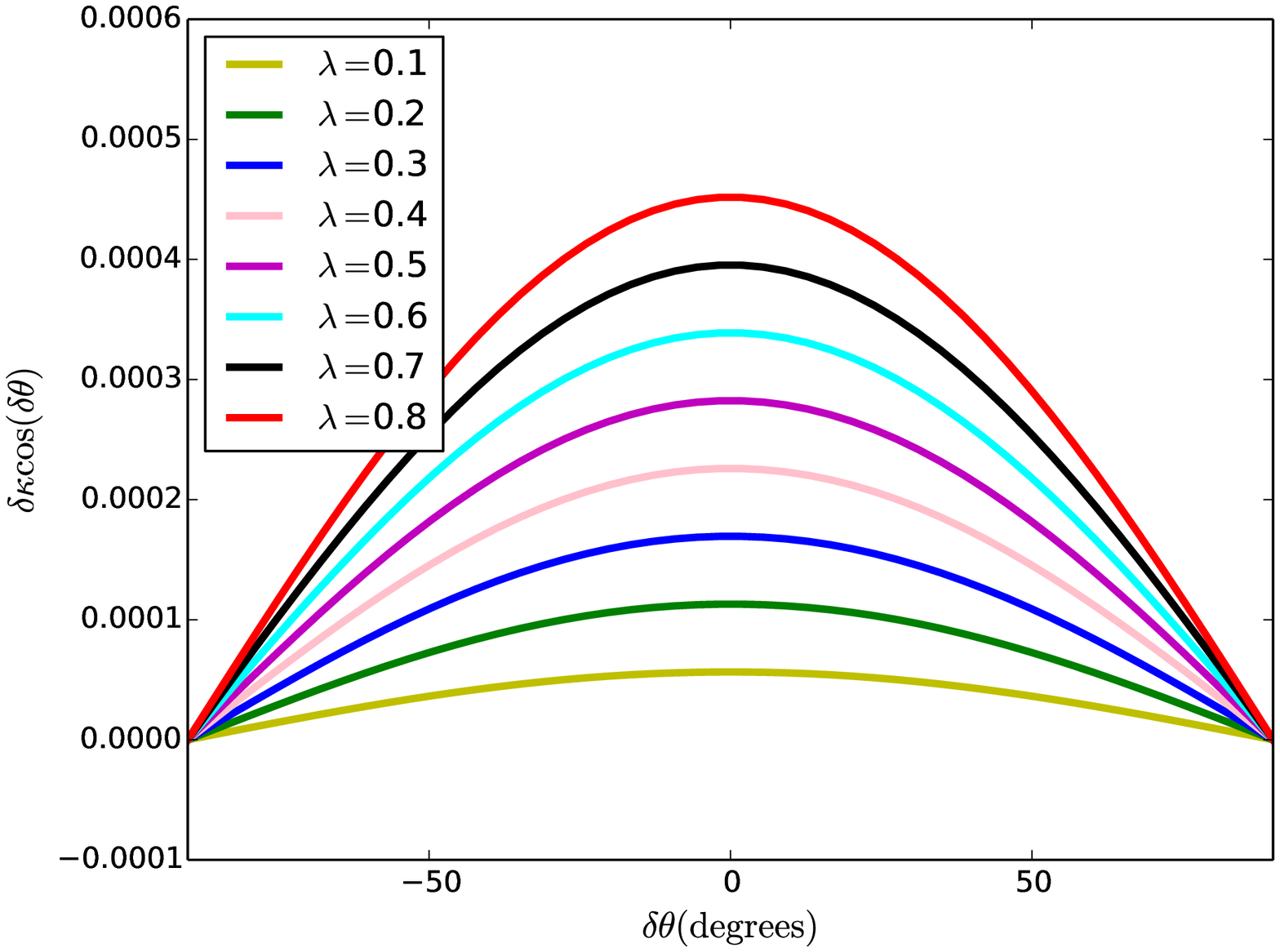}
    \caption{\textit{Upper}:$\delta\kappa$ as a function of $\delta\theta$ and $\sigma_v$ (in [km/s]) at fixed $\lambda=0.5$.
    \textit{Lower}: $\delta\kappa$ as a function of  $\delta\theta$ and $\lambda$ at fixed $\sigma_v$=1000km/s.}
    \label{fig:angdep}
\end{figure}

\begin{figure}
   %\centering
    \includegraphics[width=9cm,height=7cm]{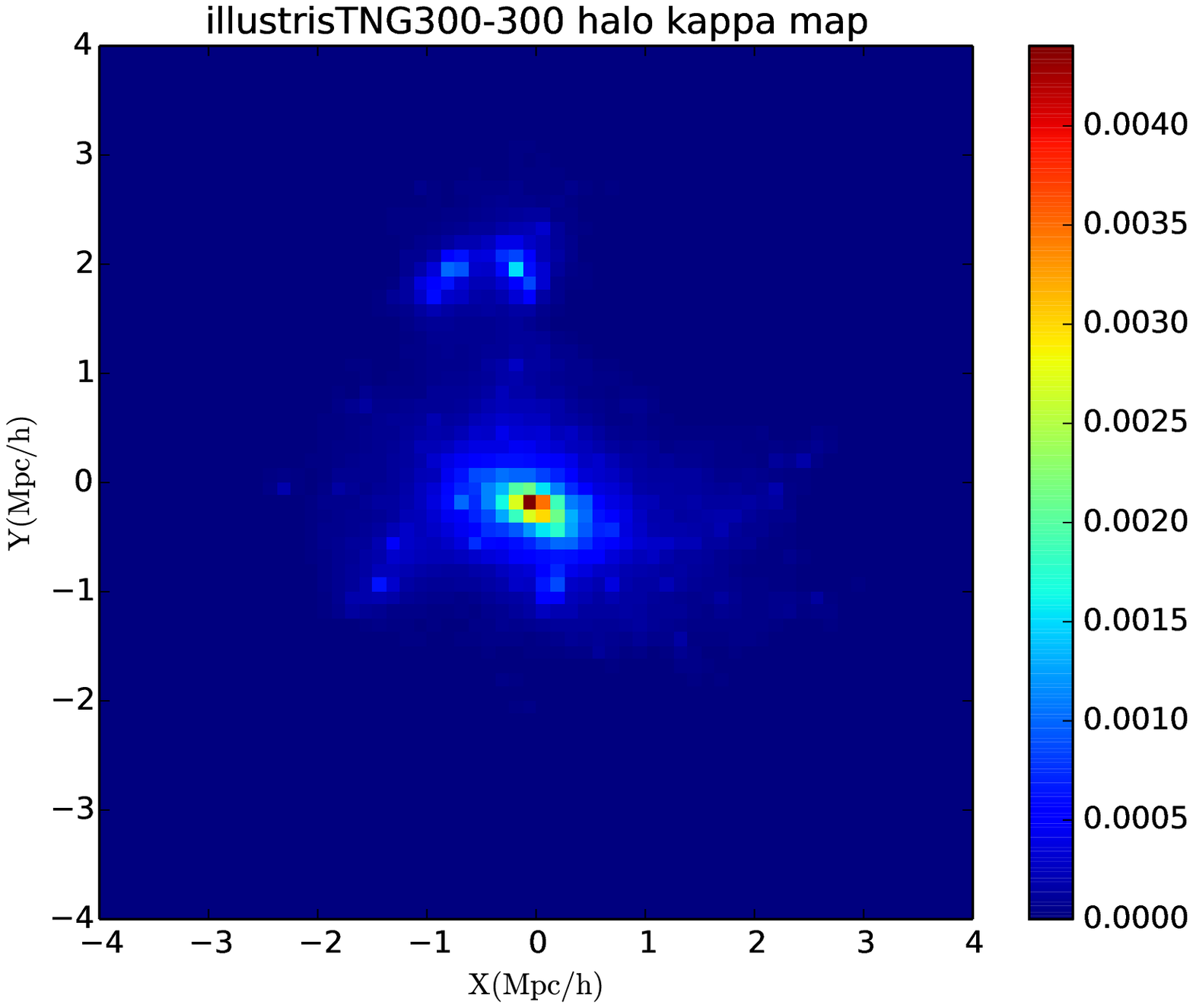}
    \includegraphics[width=8cm,height=7cm]{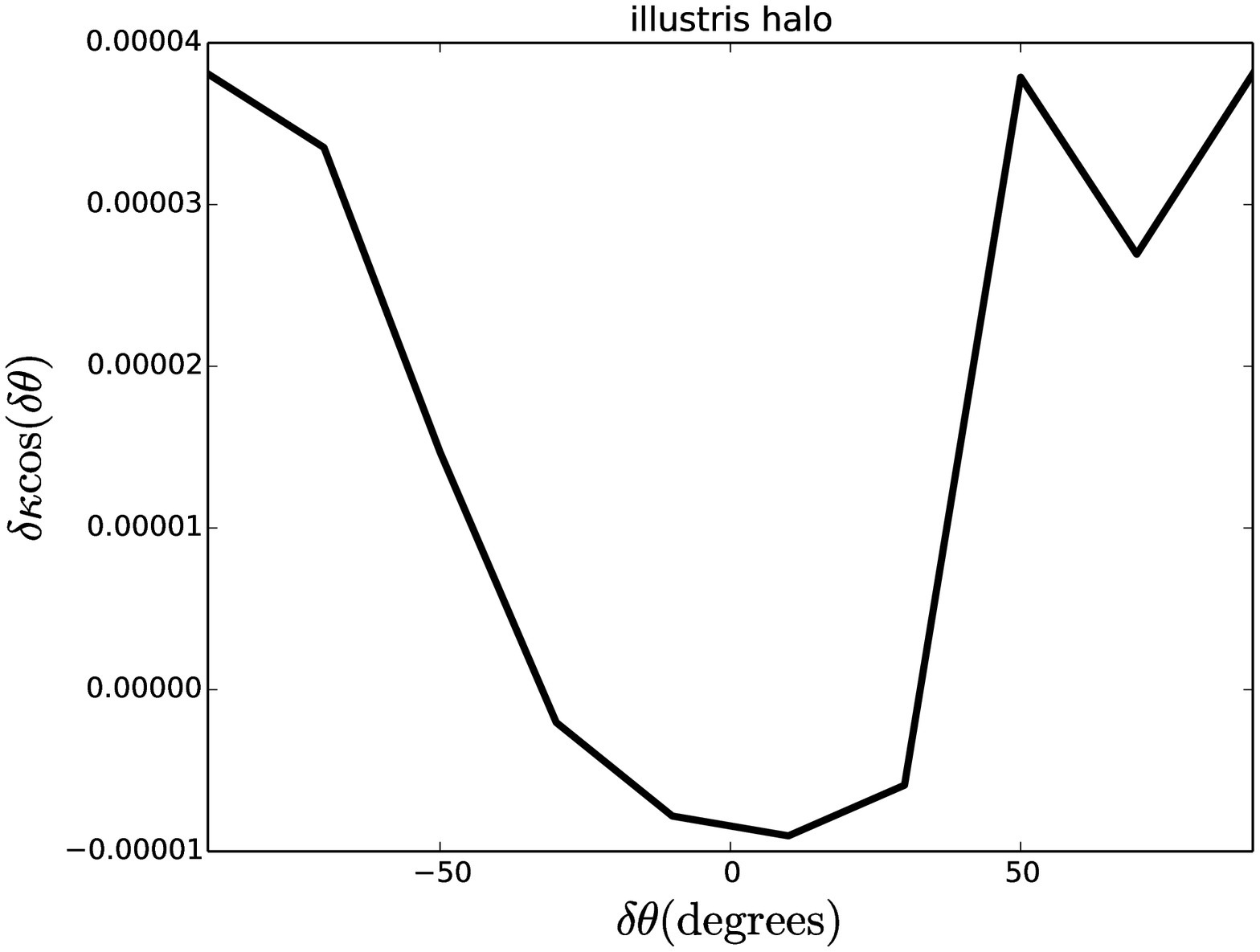}
    \caption{\textit{Upper}: $\kappa$ map of an illutstrisTNG300-300 halo, with $\log (M_h/h^{-1}M_{\odot}) \sim 15 $ at redshift $z=0.2$ and source redshift $z=0.4$. \textit{Lower}: $\delta\kappa$ measured around this halo as a function of $\delta\theta$. 
    %The maximum contribution is 0.0004. 
    The $\delta\theta$-dependence is very different than the one shown in Fig.~\ref{fig:angdep}.}
    \label{fig:illustrishalo}
\end{figure}

We show the dependence of $\delta\kappa$ on the azimuthal angle, Eq.~\eqref{eq:dkapm}, in Fig.~\ref{fig:angdep}. The angular dependence (angle between the true rotation axis and an arbitrary axis) of $\delta\kappa$ is clear. However, this is difficult to estimate by stacking without any knowledge of the rotation axis. We will show later in Sec.~\ref{sec:result} that, we can infer the rotation axis by observing the same pattern of angular dependence as in Fig.~\ref{fig:angdep}.
Such specific dependence can be used to distinguish gravitomagnetic distortions from other sub-leading lensing contributions, such as from halo asymmetry, biased tracers, and so on. 
In particular, one should keep in mind that irregularities in the halo shapes can lead to a very different signal than the one we derived here, assuming perfect spherical symmetry and a SIS density profile.
In Fig.~\ref{fig:illustrishalo}, we show for illustration the measured $\kappa$ map and values of $\delta \kappa$ as a function of $\delta \theta$ on a typical halo identified in illustrisTNG300-300~\citep{nelson2018mnr}. 
Note that we did not add shape noise here, since for a single cluster, the shape noise will be overwhelmingly dominant. %Furthermore, because no simulations sothere is no gravitomagnetic estimation from the simulation. What is shown is the estimator from Eq.~\ref{eq:deltakappa} as applied to a static case.}
The signal is comparable in size to a gravitomagnetic distortion with $\sigma_v$=1000 km/s and $\lambda$=0.2, but we can see that the halo morphology leads to a very different profile for $\delta \kappa (\delta \theta)$.
Furthermore, the effect of biased tracers has not been taken into account here (this will be investigate in an upcoming work). 
However, at this stage, we expect that biased tracers will present a similar signal as the one for halos for gravitomagnetic effect, at least for isolated clusters.
Thus, to hope detecting Einstein-Thirring-Lense effect in weak lensing surveys, one should consider relatively isolated clusters, i.e. with cylindrical selection (e.g.~\cite{mandelbaum2006}), or isolated galaxies (e.g.~\cite{luo2020}).

%We also test $\delta \kappa$ introduced by irregularity of halo shapes from illustrisTNG300-300~\citep{nelson2018mnr}. We show that halo morphology has very different curve from that of gravitomagnetic effect, and the value is comparable  with $\sigma_v$=1000 km/s and $\lambda$=0.2. We do address that we have not included the large scale structure yet, which will be studied in the upcoming work. However, we do not think the large scale structure has significant effect as GRM, because the strongest signals are from cluster regions, unless there is another cluster nearby. So in observation, one has to select relatively isolated clusters, i.e. cylindrical selection as in \cite{mandelbaum2006} or isolated galaxies as in \cite{luo2020}. 
In the following section, we perform a suite of simulations to test the model above, with realistic shape noise of galaxies and galaxy number density to evaluate the detectability in LSST-like surveys.
\\
\section{Simulation} 
\label{sec:toy}
We simulate $\kappa$ maps as described by Eq.~\eqref{eq:kappa_grm} assuming a shape noise of 0.3 and galaxy number density of 50 per sq. arcmin following the characteristics of LSST~\citep{lsst}. 
In real observations, $\kappa$ maps are reconstructed
from the shear distortion field $\gamma_i, i=1,2$. All the simulated rotating halos are assuming SIS profiles in order to be consistent with our formulations.
Thus, the error propagation of the shape noise to $\kappa$ can be estimated in Fourier space as~\citep{mrlens2011}:
\begin{equation}
    \tilde{\kappa}=\tilde{P_1}\tilde{\gamma_1}+\tilde{P_2}\tilde{\gamma_2},
\end{equation}
where $\tilde{P_1}$ and $\tilde{P_2}$ are defined as:
\begin{equation}
    \tilde{P_1} =\frac{k_1^2-k_2^2}{k_1^2+k_2^2}, \qquad \tilde{P_2} =\frac{2k_1^2k_2^2}{k_1^2+k_2^2}.
\end{equation}
$\tilde{P_1}=0$ when $k_1^2=k_2^2$ and $\tilde{P_2}=0$ when $k_i=0,(i=1,2)$.
Propagating the shape noise of both components of $\gamma_i$, $N_i$, the measured $\kappa$ is then given by:
\begin{equation}
    \tilde{\kappa}_{n}=\tilde{P_1}(\tilde{\gamma_1}+\tilde{N_1})+\tilde{P_2}(\tilde{\gamma_2}+\tilde{N_2}),
\end{equation}%\adg{fix the tilde}
such that the error on $\kappa$ is:
\begin{equation}
   \tilde{N}_{\kappa} = \tilde{P_1}\tilde{N_1}+\tilde{P_2}\tilde{N_2}.
\end{equation}
Here, we simply have $\tilde{N}_{\kappa}=N_{\gamma}=0.3$, where $N_{\gamma}$ is the shape noise, given that $N_1=N_2=0.3$ and $k_1=k_2$.
In this work we neglect the error and bias associated to the reconstruction algorithm.
The final error that we use for our simulations is then $\sigma =\frac{0.3}{\sqrt{N_{gal}}}$. 

%We will leave the reconstruction part in our next work together with illustrisTNG simulation. 
%We simulate stacked clusters given scattered observational value of $\lambda=0.1, 0.2, 0.25$, with fixed $\sigma_v=1000$. The numbers of the stacked clusters are mimicking LSST volume as we estimated later. In order to access how many clusters, and how large the $\lambda$ value can generate significant signals. We simulate 6 sets of simulat

%\subsection{Single cluster simulation}
We simulate five sets of stacked rotating clusters with fixed velocity dispersion of $1000$km/s. 
Table~\ref{tab:sims} shows the specification chosen for each simulation.
The first three (Simulations 1, 2, and 3) are with rotation parameters $\lambda=0.1,0.2,0.25$, 
These values are motivated by the following.
We choose from~\cite{miller2005AJ} the C4 identification algorithm as informative reference for cluster characteristics: a cluster with richness of 36 has been identified with a velocity dispersion $2182$km/s. 
However, we restrict ourselves to a more conservative upper limit for the velocity dispersion of $1000$km/s from the observation of ABELL 2255 identified by the C4 identification algorithm.
This corresponds to halo mass of $\log (M_h/h^{-1}M_{\odot}) \sim 15 $, where $M_{\odot}$ is a solar mass, according to the scaling relation of \cite{zahid2016}. 
%\adg{A comment: Wouldn't be interesting to also see simulations for a more optimistic case with $\sigma_v=2000$km/s? or is this completely unrealistic? Also one can suppose that the identification algorithm becomes better? What would be the threshold in terms of cluster numbers / velocity dispersion / lambda for a detection ($5 \sigma$ exclusion from $0$)? That would be interesting to see.}
%\adb{Yes, I agree. But we have tried 2000km/s and $\lambda=0.8$. The shape noise still overwhelms the signals. Nothing can be detected. So we stick to stacking technique.}
According to~\cite{vitvitska2002ApJ}, the rotation parameter is about $\langle \lambda \rangle \approx 0.05$ on average, however, faster rotating halos can be expected from a distribution such as in Eq.~\eqref{eq:plambda}. 
If we consider SDSS DR7 north cap spectroscopic survey volume ($\sim 7500$deg${}^2$ after masking) within the redshift range $\left[0.01,0.2\right]$, there are 26 clusters with halo mass $>10^{15}h^{-1}M_\odot$. 
If we consider LSST-like survey with $\sim 20000$deg${}^2$ and a redshift range from 0.01 to 1, the volume is roughly 184 times the SDSS DR7 volume group catalog. This estimation is based on a recent group finder algorithm by \cite{yang2020arXiv}, who modify the halo-based group finder designed for SDSS survey to adopt the photometric only data for DESI Legacy Survey~\citep{DESI2019AJD}. The group catalog has photometric redshift ranging from 0.0 to 1.0.
In that case, roughly about 4785 clusters with $\log (M_h/h^{-1}M_{\odot}) \sim 15 $ can be detected. 
For $\lambda>0.1$, only $\sim 400$ clusters can be detected, with mean $\lambda=0.128$.
The number of detectable clusters further reduces to 13 for $\lambda>0.2$ and to 3 for $\lambda>0.25$.  

\begin{figure}
   %\centering
    \includegraphics[width=9cm,height=7cm]{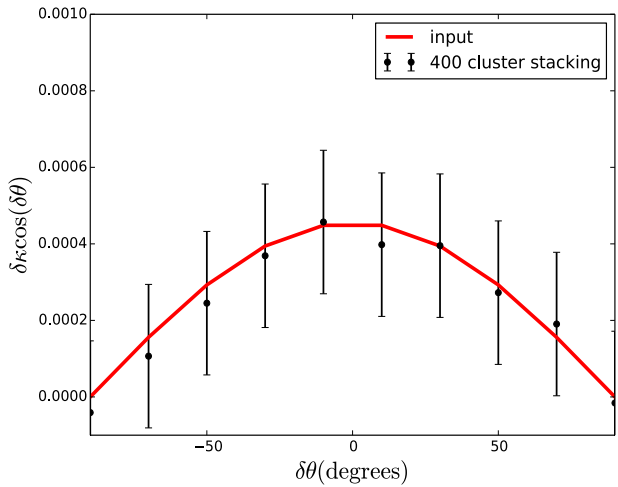}
    \caption{The angular dependence of $\delta\kappa$ from 400 stacked clusters with $\sigma_v=1000$km/s and $\lambda=0.8$. The black dots are the mocked data. The theoretical input is shown by the red line.} 
    \label{fig:noisy}
\end{figure}

We also simulate six unrealistic cases with $\lambda=0.3-0.8$ and velocity dispersion of 1000 km/s and 400 stacked clusters with zero scatter between the rotation axis indicator and the real rotation axis (Simulation 4). Fig.~\ref{fig:noisy} shows one of the simulations for the 4th set with $\lambda=0.8$ assuming zero scatter of on rotation axis.
The last simulation (Simulation 5) is set to estimate the signal reduction from the misalignment between the true rotation axis and the tracer axis from observations. 
%Table~\ref{tab:sims} generalizes the characters of each simulation to probe the detectability of $\delta\kappa$ for real observations.
 
\begin{table}[ht]
\caption{5 suites of simulations used in this work with fixed velocity dispersion $\sigma_v=1000$km/s, various rotation parameters $\lambda$ and various scatters $\sigma_{\theta}$ of misalignement $\delta\theta$, and number of simulated cluster halos.}
\begin{center}
\begin{tabular}{ccccc}
    \hline
    Simulation & $\sigma_v$ (km/s) & $\lambda$ & $\sigma_{\theta}$(deg)  & Num\\
    \hline
    Sim 1 &1000 & $>0.1$   & 0.0  & 412\\
    \hline
    Sim 2 &1000 & $>0.2$   & 0.0  & 13\\
    \hline
    Sim 3 &1000 & $>0.25$ & 0.0   & 3 \\
    \hline
    Sim 4 &1000  & 0.3-0.8   & 0.0   & 400\\
    \hline
    Sim 5 &1000  & 0.8    & [1,5,10,20,30]  & 400\\
    \hline
\end{tabular}
\end{center}
\label{tab:sims}
\end{table}

\section{Results}
\label{sec:result}
\subsection{Simulation results}
\begin{figure}
   %\centering
    \includegraphics[width=9cm,height=7cm]{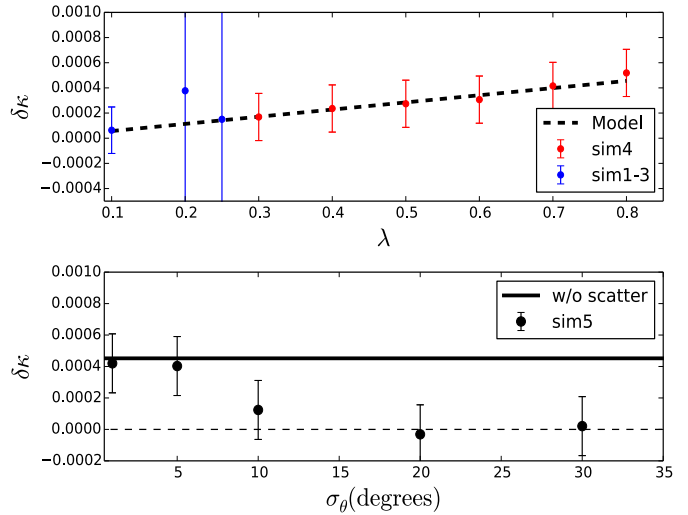}
    \caption{
    Measurements of $\delta \kappa$ on simulations.
    \textit{Upper}: Simulations 1 to 3 (in blue), simulation 4 (in red). 
    The theoretical input is depicted by the dashed black line. 
    \textit{Lower}: Simulation 5 for various scatters $\sigma_{\theta}$ %\adg{should this be $\delta \theta$ or is this something else?}
    representing the misalignment of the true halo rotation axis with the tracer axis. 
    The theoretical input without scatter is shown by the continuous black line.
    %causes the suppression of $\delta\kappa$.
    } 
    \label{fig:results}
\end{figure}

We now investigate the detectability of gravitomagnetic distortion in weak lensing surveys by measuring $\delta\kappa$ on the simulations designed in previous section.
The results are shown in Fig.~\ref{fig:results}.

Let us first focus on Simulations 1, 2, and 3, designed to mock LSST-like surveys: for $\lambda>0.1$, we have 412 cluster halos of mass $>10^{15}h^{-1}M_\odot$, for an average of $\lambda$ around 0.128. 
Only 13 clusters are found for $\lambda>0.2$, and 3 for $\lambda>0.25$. 
This is summarized in Table~\ref{tab:sims}.
As expected, we find that $\delta \kappa$ is smaller for Simulation 1 than for Simulations 2 and 3, as on average the rotation of halos is smaller. 
Nevertheless, as the number of stacked halos in Simulation 1 is bigger, the error bars are much smaller. 
For the other two, which only has 13 and 3 clusters, the shape noise overwhelms the signal by a factor of $\sim 10$.
In all these `realistic' cases, we find null detection of gravitomagnetic effect, given that $\delta \kappa$ is compatible with zero within $1\sigma$.

%The smaller the sample size, the larger the noise. This can be seen from the upper panel of Fig.~\ref{fig:results}.

%The blue data points from the upper panel of Fig.~\ref{fig:results} show the simulated results from simulation 1 to 3, assuming clusters with velocity dispersion of 1000 km/s. The differences among those three are in two folds, $\lambda$ and number of stacked clusters. The noise of the simulation 1 is the lowest due to the stacking number, its value is nonetheless the smallest. For the other two, which only has 13 and 3 clusters, the shape noise overwhelms the signal by a factor of 10.0. 

We perform further checks with Simulations 4 and 5. 
Simulation 4 is set to understand the relation between the measured $\delta\kappa$ and $\lambda$. The values of $\lambda$ ranges from 0.3 to 0.8, for fixed number of 400 of stacked clusters. 
The theoretical prediction, Eq.~\eqref{eq:dkap}, agrees with Simulation 4 data points. 

Simulation 5 is designed to study the effect of a misalignment between the true rotation axis of clusters and the ones selected using observational tracers, such as the major axis of central galaxies~\citep{okumura2009ApJ}, the distribution of satellite galaxies inside clusters~\citep{baxter2016MNRAS}, or spin axis of spiral galaxies~\citep{zhang2015}. 
In general, the tracers are misaligned with their dark counterpart. One famous case is the bullet cluster~\citep{cloawe2006ApJ}, in which the baryonic distribution significantly differs the dark matter distribution inferred from weak lensing. 
To probe such discrepancy, we choose randomly for each cluster in Simulation 5 the misalignment angle between the tracer axis and the true rotation axis among a normal distribution centered on 0 with scatter: $\sigma_{\theta}$ = 1, 5, 10, 20 and 30 degrees. 
We find that a 5 degrees scatter leads to a damping of $\sim 10\%$, that remains consistent with the true input value denoted by the solid black line in the lower panel of Fig.~\ref{fig:results}. 
%For scatters larger than 20 degrees 
For larger scatters, the signal becomes too low to be detected, where even with 10 degrees scatter the signal is already consistent with zero.
This highlights the importance to select precisely the tracer axis in order to extract the gravitomagnetic distortion signal. 

\subsection{Observational indications}
We now proceed with a rough quantification of the rotation axis and speed from the groups of galaxies of SDSS DR7 using redshift selection following~\citet{yang2007apj}. 
The same method to measure $\delta\kappa(\delta\theta)$ can be used to measure the mean redshift of member galaxies separated by an arbitrary axis that has an angle of $\delta\theta$ with respect to the true axis in two dimensional projected plane. 
Here, we use $\phi$ to denote the angle between the true rotation axis and the arbitrary axis from observations in order to distinguish it from $\delta\theta$ used for simulations. The difference of the mean redshift from the two sides divided by the 2D rotation axis is given by: $\Delta z=\langle z_1\rangle-\langle z_2\rangle$, where the $1,2$ indicate the two
sides of the axis. 
Following~\citet{sofue2013pasj} in which the rotation curve of our own galaxy is found to be well fitted by a sine function, we fit $\Delta z$ with the parametrization:
\begin{equation}
    \Delta z(\phi)=z_{off}\pm z_{amp}\times \sin(\phi-\phi_0).
\label{eq:curve}
\end{equation}
The first term on the right hand side of the Eq.~\eqref{eq:curve} is the offset of the group along the line of sight, which should be zero since we choose the central galaxies as the reference one such that the peculiar velocity of the cluster
becomes zero~\citep{hwang2007apj}. 
However, we caution that there might be uncertainties in the choice of the Brightest Central Galaxies (BCG).
For our purpose, we consider that a nonzero offset will be much smaller than the error and thus can be neglected. 
The amplitude $z_{amp}$ represents the rotation speed along the line of sight, and $\phi_0$ is the rotation axis angle with respect to the East to North. 
The relation between $\Delta z$ and velocity dispersion along the line of sight are related by~\citep{danese1980aa}:
\begin{equation}\label{eq:vlos}
    v_{los}=c\frac{\Delta z}{1+z_{bcg}},
\end{equation}
where $z_{bcg}$ is the redshift of the BCG.
With this procedure outlined above, we are thus able to extract all information to evaluate $\delta \kappa$.

\begin{figure*}
    \centering
    \includegraphics[width=7cm,height=5.25cm]{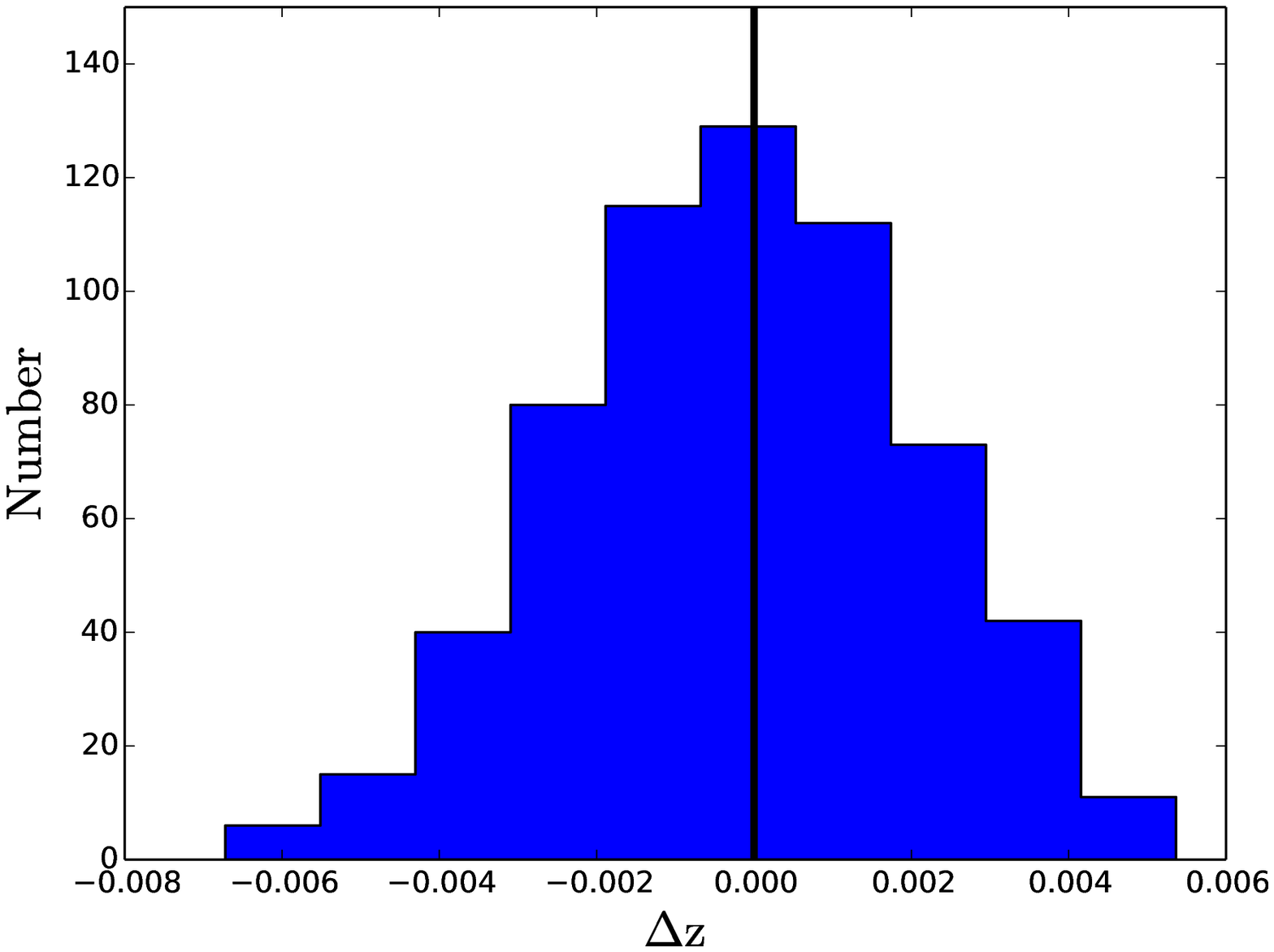}
    \includegraphics[width=7cm,height=5.25cm]{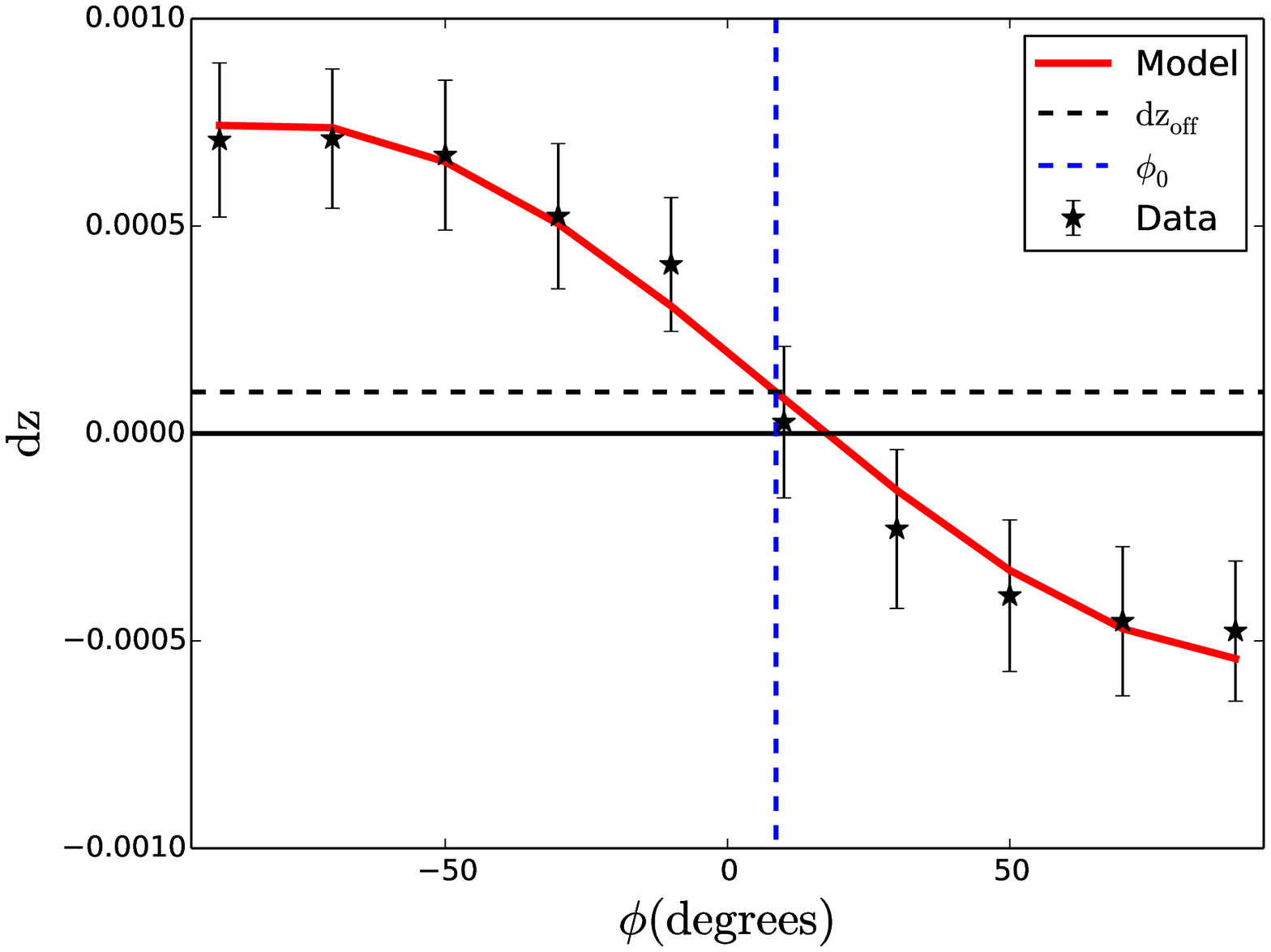}
    \includegraphics[width=7cm,height=6.125cm]{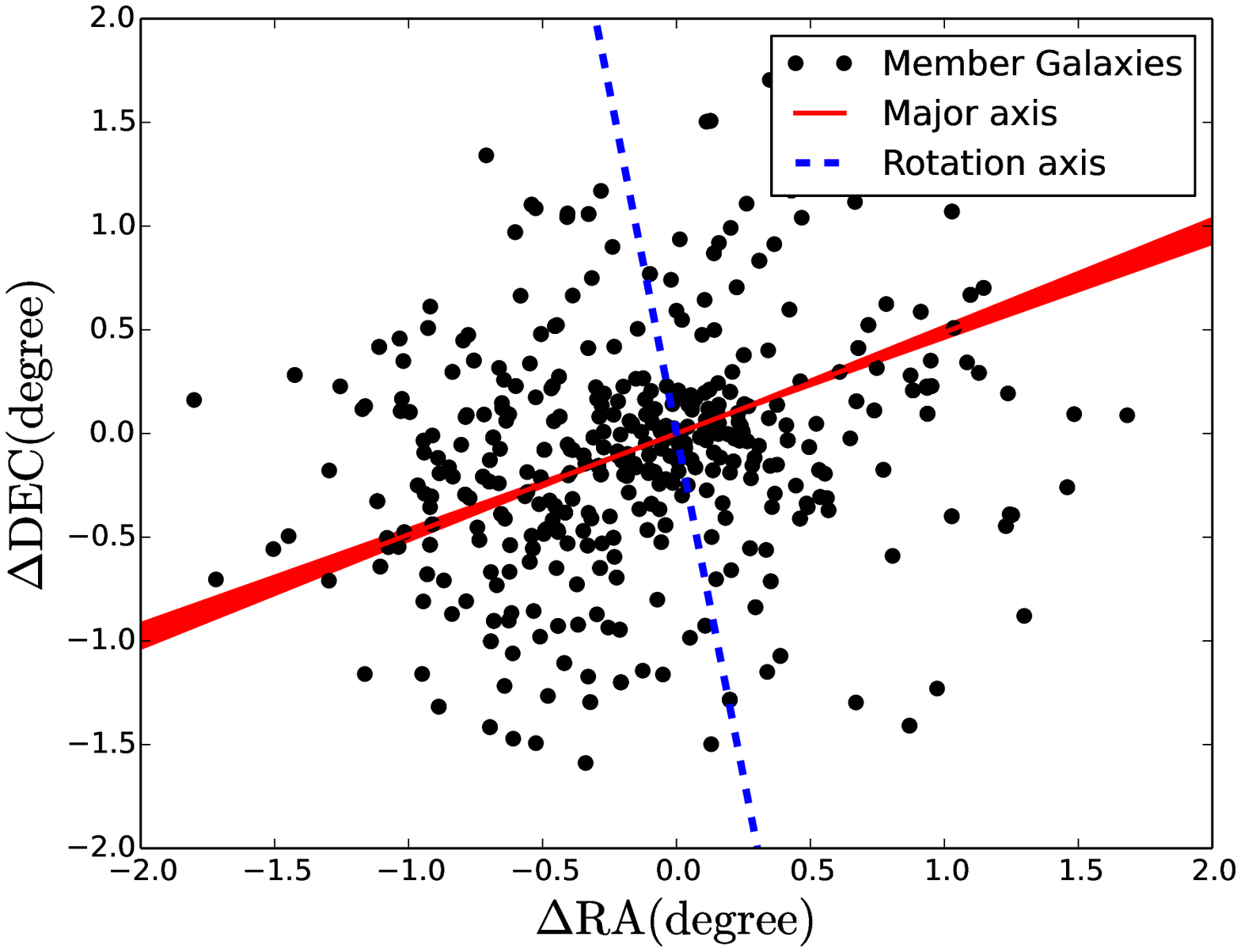}
    \caption{\textit{Upper left:} The distribution of $\Delta z$ of the richest group in Y07 catalog with scatter of 0.0022 corresponding to roughly 667.8km/s in terms of velocity dispersion. The halo mass of this cluster from abundance matching is $\log (M_h/h^{-1}M_{\odot}) \sim 15 $. 
    \textit{Upper right:} The rotation curve fitted with a sine function as in Eq.~\eqref{eq:curve}. This provides indication of the rotation axis alignment. The amplitude infers a line-of-sight rotation speed of 195.0km/s. The $\lambda$ parameter is then equals 0.292. 
    \textit{Lower:} The major axis of satellite distribution denoted as red solid line has 72.62 degree offset from the rotation axis denoted as the blue dashed line. The errors of the major axis estimated from bootstrap is denoted as the red band, which is 1.42 dergrees. }
    \label{fig:group1}
\end{figure*}

As an illustration, we select the richest group in~\cite{yang2007apj} catalog with 623 members with velocity dispersion of about 667.8km/s estimated from the scatter of $\Delta z= z_{member} - z_{bcg}$ as shown in Fig.~\ref{fig:group1}, and with line-of-sight rotation speed of about 195.0km/s based on Eq.~\eqref{eq:curve} and Eq.~\eqref{eq:vlos}. 
The right upper panel of Fig.~\ref{fig:group1} illustrates that the rotation axis alignment can be measured by fitting the rotation curve using a sine function as in Eq.~\eqref{eq:curve}. We emphasize that, however, one also has to account for the projection effects in order to obtain a reliable measurement of the rotation axis. This will be discussed in a followup study based on the combination of observational group catalog and simulations.
This leads to $\lambda=0.292$, which is extremely unlikely given the probability distribution $P(\lambda)$ given in Eq.~\eqref{eq:plambda} that we used for our simulations: the fraction of clusters with halo mass larger than $10^{15}h^{-1}M_{\odot}$ in SDSS DR7 is about $0.00006$ and the probability of a halo with $\lambda>=0.3$ is about 0.0002, yielding a joint probability of such cluster to appear in this catalog of only $1.2\times 10^{-8}$, according to the probability distribution inferred from simulations~\citep{vitvitska2002ApJ}. 
Yet, we note that this is still more likely than the bullet cluster. The probability of forming a bullet cluster liked object with such high colliding speed in LCDM framework is about $3.3\times 10^{-11}$ \citep{lee2010ApJ}.  
The reason of such occurence may be due to the fact that the velocity dispersion of this cluster in underestimated because SDSS spectroscopic objects need to be brighter than 17.77 mag in the r band, which means that fainter members are not included. Even if we set the velocity dispersion to be 1000km/s, $\lambda$ is still about 0.2. It thus seems that highly rotating clusters are not as rare as predicted by~\citet{vitvitska2002ApJ}.

The rotation speed is the most important quantity to estimate gravitomagnetic distortion. 
If we assume that the rotation speed is 195.0km/s, for a cluster halo mass of $\log (M_h/h^{-1}M_{\odot}) \sim 15 $, this leads to a value of $\delta\kappa=0.0002$. 
But for such
a cluster, the shape noise will be of order 0.003 which is more than 10 times larger than the signal.
\citet{manopoulou2017mnr} have already started this research on the angular dependence of the line-of-sight rotation velocity and rotation axis using Monte-Carlo mocks of rotating clusters. They found that using their algorithm up to 28\% clusters can be identified as rotating.

The error bars from Fig.~\ref{fig:group1} is estimated by bootstrap resampling method: we create 600 samples out of 623 member galaxies and take the distribution of mean value of each sample to estimate the velocity and error. 
%We proceed with this study of the angular variation of the other groups in another work to study further how to identify the rotation axis and its scaling relation with other properties. 
The lower panel of Fig.~\ref{fig:group1} shows a \textbf{$72.62\pm 1.42$ degrees} offset between the major axis of satellite distribution (red solid line) and rotation axis (blue dashed line) which is simply $\pi/2+\phi_0$. 
This means that the major axis from satellites can not be used as rotation axis and tends to be anti-aligned with the major axis in this cluster. 
Note that we mention the major axis of clusters, as it is of more interests than the minor axis in the sense that it is more likely aligned with large-scale filament structure.
The alignment between the two axis is another interesting study, which we will explore in the future. 

In this figure we have all the information needed to extract
$\delta\kappa$ from observational data, i.e. the velocity dispersion, $\lambda=v_{los}/\sigma_{v}$, and the rotation axis. 
We will further explore if this can be achieved using the group catalog from~\cite{yang2007apj} and the SDSS DR7 shape catalog of~\cite{luo2017apj} (Tang et al. in prep.). 
%Another interesting issue is whether the $p(\lambda)$ distribution from observations and compare that with $\Lambda$CDMsimulations. 

%In general, this effect is so subtle for weak lensing to detect. Still, the fact that there are 28\% cluster with detectable rotation.  Even for Simulation 1, the signal is much smaller than the noise level without taking the scatter into account. In order to get 1 $\sigma$ detection in simulation 1 liked condition, one needs to stack at least about 3600 clusters with velocity dispersion of 1000 km/s, and $\lambda>0.1$.  The shape noise is the dominate obstacle in this measurement. This may not be the case for CMB lensing map \citep{vanengelen2014apj}.  However, all those 3600 clusters must have reliable axis tracers within at most 5 degrees scatter with respect to the true axis. 

\section{Conclusion}
\label{sec:conclusion}

We have analyzed the Einstein-Thirring-Lense effect in lensing convergence maps around rotating halos. 
Making the bridge between observations and  previous theoretical works~\citep{ciufolini2003,sereno2005mnr,sereno2007}, we have constructed an estimator to measure the anistropic signal from  gravitomagnetic effects induced by the rotation of foregrounds in weak lensing surveys.
We find that this signal is two or three orders of magnitude smaller than the distortion from the static halo potential, in accordance with~\cite{lazaro2018mnras}.

Assuming lenses with a spherical SIS density profile, we have run a suite of simulations calibrated for LSST-like survey, for which we estimate that only about 400 clusters with relatively high rotation parameters $\lambda>0.1$ can be found. 
Applying our estimator on the simulations,  even by stacking $\sim400$ rotating clusters with $\lambda>0.1$ and velocity dispersion within the clusters of $\sigma_v>1000$km/s, assuming perfect knowledge of the halo rotation axis.
We also simulate data to test observational effects, such as the impact on the signal amplitude of a misalignment between the observational tracer axis and the halo rotation axis, finding that the signal is strongly reduced for a misalignment scatter of more than $5$ degrees.

However, we observe that the chosen characteristics for our mocks, in particular the dependence on the rotation parameter $\lambda$, is strongly motivated based on inputs from N-body simulations.  
In real observations, fast rotating clusters may not be as hard to find as in simulations. 
Indeed, we have further discussed methods to identify rotating clusters in maps from spectroscopic surveys. By selecting the most massive cluster from~\cite{yang2007apj}, we have found that this cluster has an extremely big angular momentum with $\lambda=0.295$, which is very unlikely given the distribution we assumed for $\lambda$ based on simulations.
This serves as a preliminary indication that the story might be different in real observations: rotating halos may be more abundant and at higher speed in our universe than what simulations show, motivating to investigate further gravitomagnetic effects in the context of weak lensing. This will be explored in an upcoming work.

We finish by highlighting a number of interesting open questions  related to the current work. 
\begin{itemize}
    \item We found that the major axis of Satellites position distribution in clusters can not be used to probe gravitomagnetic effect, as it is completely misaligned with the cluster rotation axis. However, the alignment between the major axis and the cluster rotation axis can be an interesting topic to study, as it can hint on the direction of the real rotation axis.
    \item We have sketched a procedure how to select rotating clusters in group catalog. This can be used to measure the distribution of $\lambda$, to see whether it is consistent with the one predicted from first principles or with the use of simulations.
    \item Modified gravity may lead to different lensing signals, in particular through the gravitomagnetic potential. Although gravitomagnetic distortions are difficult to detect, it may be interesting to study the impact of modification to gravity on frame-dragging effects, as the resulting signal may be different in shape and size.
    \item The elongation of the dark matter halo along the major axis does not introduce any signal to $\delta\kappa$ due to the subtractive nature of the estimator. As long as the density profile is symmetric with respect to the rotation axis, this elongation will not cause any extra $\delta\kappa$.
\end{itemize}
We leave these issues to future studies.

\section*{Acknowledgement}
We are grateful to X. Kong, Z. F. Sheng, D. Nagai, J. J. Zhang, A. Sen, W. Hossain and B. Dinda for valuable discussions. We have also received valuable suggestions and discussions from Marco Bruni at University of Portsmouth, Baojiu Li and Cristian Barrera-hinojosa from Durham Unviversity and we are very grateful for that.
This work is supported in part by the NSFC (Nos. 11653002, 1201101448, 11961131007, 11722327, 11421303), by the CAST(2016QNRC001), by the National Youth Talents Program of China, by the Fundamental Research Funds for Central Universities, and by the USTC Fellowship for International Cooperation. WL and SP are supported by the World Premier International Research Center Initiative (WPI Initiative), MEXT, Japan. SP is supported partially by the National Key Research and Development Program of China Grant No.2020YFC2201502, and by JSPS Grant-in-Aid for Early-Career Scientist No. 20K14461. All numerics were operated on the computer clusters {\it LINDA} \& {\it JUDY} in the particle cosmology group at USTC and {\it gfarm} cluster at Kavli IPMU, the University of Tokyo.

\newpage
%\bibliographystyle{apsrev4-1}
%\bibliography{cgnbody}
\bibliographystyle{apj}
\pagestyle{plain}
\bibliography{grmkappa}
\end{document}